\documentclass[twocolumn,showpacs,preprintnumbers,amsmath,amssymb]{revtex4}
\usepackage{amssymb}
\usepackage{amsmath}
\usepackage{graphicx}
\usepackage{epsfig}

\begin{document}

\title{Superconducting decay length in a ferromagnetic metal}

\author{D.\,Yu.~Gusakova}
\email{dariamessage@yandex.ru} \affiliation{Nuclear Physics
Institute, Moscow State University, Moscow, 119992 Russia}

\author{A.\,A.~Golubov}
\affiliation{Faculty of Science and Technology, University of
Twente, 7500 AE Enschede, The Netherlands}

\author{M.\,Yu.~Kupriyanov}
\affiliation{Nuclear Physics Institute, Moscow State University,
Moscow, 119992 Russia}

\begin{abstract}
The complex decay length $\xi $ characterizing penetration of
superconducting correlations into a ferromagnet due to the proximity
effect is studied theoretically in the frame of the linearized
Eilenberger equations. The real part $\xi_1 $ and imaginary part
$\xi_2 $ of the decay length are calculated as functions of exchange
energy and the rates of ordinary, spin flip and spin orbit
electronic scattering in a ferromagnet. The lengths $\xi_{1,2}$
determine the spatial scales of, respectively, decay and oscillation
of a critical current in SFS Josephson junctions in the limit of
large distance between superconducting electrodes. The developed
theory provides the criteria of applicability of the expressions for
$\xi _{1}$ and $\xi_{2}$ in the dirty and the clean limits which are
commonly used in the analysis of SF hybrid structures.
\end{abstract}

 \maketitle

The decay length $\xi$ is an important material parameter which
characterizes the scale of penetration of superconducting correlation into a
non-superconducting material across an interface with a superconductor. The
critical current $I_{C}$ in a Josephson junction scales exponentially with
the distance between the superconducting electrodes $L$ if $L$ is larger
that $\xi $: $I_{C}\varpropto \exp \left\{ -L/\xi \right\}.$ In nonmagnetic
materials the decay length is a real number, while in ferromagnets $\xi $ is
a complex number (see \cite{Likharev}-\cite{bverev} for the reviews). In
particular, if the condition of so-called dirty limit is fulfilled in the F
metal, the decay length is
\begin{equation}
\xi ^{-1}=\xi _{1}^{-1}+i\xi _{2}^{-1},~\xi _{1,2}^{-1}=\sqrt{\frac{\sqrt{%
\left( \pi T\right) ^{2}+H^{2}}\pm \pi T}{D_{F}}},  \label{ksi_d}
\end{equation}%
where $D_{F}$ and $H$ are the diffusive coefficient and the exchange field
in a ferromagnet, respectively. In the clean limit
\begin{equation}
\xi _{1}^{-1}=\xi _{0}^{-1}+\ell ^{-1},~\xi _{0}^{-1}=\frac{2\pi T}{v_{F}}%
,~\xi _{2}^{-1}=\xi _{H}^{-1}=\frac{2H}{v_{F}},  \label{ksi_c}
\end{equation}%
where $v_{F}$ is the Fermi velocity in a ferromagnet and $\ell $ is the
electron mean free path. From (\ref{ksi_d}), (\ref{ksi_c}) it is clearly
seen that for dirty materials $\xi _{2}>\xi _{1}$, and in the limit of large
$H>>\pi T$, the characteristic lengths are nearly equal $\xi _{1}\approx \xi
_{2}.$ In the clean limit these length scales $\xi _{1}$ and $\xi _{2}$ are
completely independent.

The existing experimental data obtained up to now in SFS Josephson junctions%
\cite{ryazanov2001}-\cite{Born} can be separated into two groups depending
on whether weak or strong ferromagnet was used for junction fabrication. To
be considered as a weak ferromagnet, the dilute ferromagnetic alloys (e.g. Cu%
$_{1-x}$ Ni$_{x},$) should be in the range of concentration close to the
critical one $(x\approx 0,5)$. The electron mean free path in these alloys
is very small providing the fulfillment of the dirty limit conditions. As a
result, the observed relation between the decay $\left( \xi _{1}\right) $
and oscillation $\left( \xi _{2}\right) $ lengths $\xi _{2}\gtrsim \xi _{1},$
is close to that following from (\ref{ksi_d}). It is necessary to point out
that in some experiments \cite{ryazanov1} the observed difference between $%
\xi _{2}\ $and $\xi _{1}$ is so large that it can not be explained by
temperature factor in (\ref{ksi_d}) only and spin-dependent scattering
processes should be taken into account \cite{ryazanov1},\cite{Fuare}.

Contrary to that, in the structures with strong ferromagnet\cite{Shelukhin},
\cite{Born} (Ni, Ni$_{3}$Al), the relation between $\xi _{1}$ and $\xi _{2}$
is just the opposite and large ratio $\xi _{1}/\xi _{2}\sim 10$ was observed
in Ni$_{3}$Al \cite{Born}. Therefore more complex model should be developed
for the data interpretation.

Most of previous theoretical work on SF hybrids was performed
assuming the dirty limit (see \cite{GKI}-\cite{bverev}), and only
first order corrections to the decay length in small parameter $l\xi
_{H}\ll 1$ were discussed in \cite{Buzdin, Tagirov}. Properties of
SF structures in the clean limit were also studied in a number of
papers, see e.g. \cite{B1,Demler,Volkov,Volkov1}. The purpose of this work is
to develop general theory describing the decay length $ \xi $ in a
ferromagnet for any relation between $\xi _{0}$, $\xi _{H}$ and
$\ell $.

To do this we consider a generic SFS Josephson junction with arbitrary
transparency of SF interfaces and large thickness of the F layer $L\gg \xi
_{1}.$ It is well known\cite{Likharev}-\cite{bverev} that the critical
current of this structure should fall exponentially with $L$%
\begin{equation*}
I_{C}=I_{0}\exp \left\{ -L/\xi \right\} .
\end{equation*}%
Here the prefactor $I_{0}$ depends on physical properties of SF
interfaces and the nearby S and F regions, while $\xi $ depends only
on bulk parameters of F material and can be
obtained\cite{Kupclean},\cite{KupLuk} as the solution of linearized
quasiclassical Eilenberger equations \cite{eilenberger}. These
equations are valid at the distances from the interfaces larger then
$\xi $ and have the form
\cite{eilenberger},\cite{BuzRev},\cite{bverev}

\begin{equation*}
(\xi _{0}^{-1}\pm i\xi _{H}^{-1})f_{\pm }+\cos \theta \frac{\partial }{%
\partial x}f_{\pm }=
\end{equation*}

\begin{equation}
=\ell _{eff}^{-1}\left( \left\langle f_{\pm }\right\rangle -f_{\pm }\right)
+\ell _{soeff}^{-1}\left( f_{\mp }-f_{\pm }\right) ,  \label{f}
\end{equation}

\begin{equation}
\ell _{eff}^{-1}=\ell ^{-1}+\ell _{z}^{-1}+2\ell _{x}^{-1},  \label{Leff}
\end{equation}%
\begin{equation}
\ell _{soeff}^{-1}=\ell _{so}^{-1}-\ell _{x}^{-1},~\left\langle
...\right\rangle =\int_{0}^{\pi }\left( ...\right) \sin \theta
d\theta . \label{Lsoeff}
\end{equation}%
Here $\theta $ is the angle between the direction of electron velocity $%
v_{F} $ and the $x-$axis, which is oriented perpendicular to the interfaces,
$f_{\pm }=f_{\pm }(x,\theta )$ are the quasiclassical Eilenberger functions
describing the behavior of spin up and spin down electrons in the presence
of exchange field $H$ oriented parallel to the SF\ interfaces. The
parameters $\ell _{so}=v_{F}\tau _{so},\ell _{z}=v_{F}\tau _{z}$ $\ell
_{x}=v_{F}\tau _{x},$ are the electron mean free paths for parallel and
perpendicular to the direction of $H$ magnetic scattering, while $\ell
_{so}=v_{F}\tau _{so}$ is the electron mean free path for spin orbit
interaction.

Solution of Eq.(\ref{f}) has the form
\begin{equation}
f_{\pm }(x,\theta )=C_{\pm }(\theta )\exp \left\{ -\frac{x}{\xi }\right\}
,~\xi ^{-1}=\xi _{1}^{-1}+i\xi _{2}^{-1}  \label{Solution}
\end{equation}%
where $\xi$ is the effective decay length independent on $\theta $.
Substitution of (\ref{Solution}) into (\ref{f}) provides the system of two
equations for $C_{\pm }(\theta )$
\begin{equation*}
(\xi _{0}^{-1}+i\xi _{H}^{-1})C_{+}(\theta )-\xi ^{-1}\cos \theta
C_{+}(\theta )=
\end{equation*}
\begin{equation}
=\ell _{eff}^{-1}\left( \left\langle C_{+}(\theta )\right\rangle
-C_{+}(\theta )\right) +\ell _{soeff}^{-1}\left( C_{-}(\theta )-C_{+}(\theta
)\right)  \label{Cplus}
\end{equation}%
\begin{equation*}
(\xi _{0}^{-1}-i\xi _{H}^{-1})C_{-}(\theta )-\xi ^{-1}\cos \theta
C_{-}(\theta )=
\end{equation*}%
\begin{equation}
=\ell _{eff}^{-1}\left( \left\langle C_{-}(\theta )\right\rangle
-C_{-}(\theta )\right) +\ell _{soeff}^{-1}\left( C_{+}(\theta )-C_{-}(\theta
)\right)  \label{Cminus}
\end{equation}%
Solution of these equations has the form
\begin{equation}
C_{+}(\theta )=\frac{\left\langle C_{+}(\theta )\right\rangle \Lambda
_{-}^{-1}+\ell _{soeff}^{-1}\left\langle C_{-}(\theta )\right\rangle }{\ell
_{eff}\left( \Lambda _{+}^{-1}\Lambda _{-}^{-1}-\ell _{soeff}^{-2}\right) }
\label{CplusS}
\end{equation}%
\begin{equation}
C_{-}(\theta )=\frac{\left\langle C_{-}(\theta )\right\rangle \Lambda
_{+}^{-1}+\ell _{soeff}^{-1}\left\langle C_{+}(\theta )\right\rangle }{\ell
_{eff}\left( \Lambda _{+}^{-1}\Lambda _{-}^{-1}-\ell _{soeff}^{-2}\right) }
\label{CminusS}
\end{equation}%
\begin{equation*}
\xi _{10}^{-1}=\xi _{0}^{-1}+\ell _{eff}^{-1}+\ell _{soeff}^{-1}
\end{equation*}%
\begin{equation*}
\Lambda _{\pm }^{-1}=\xi _{10}^{-1}-\xi ^{-1}\cos \theta \pm i\xi _{H}^{-1}
\end{equation*}

Averaging in (\ref{CplusS}), (\ref{CminusS}) over angle $\theta $ we get the
system of two equations for $\left\langle C_{\pm }(\theta )\right\rangle .$
Its compatibility condition results in the equation for the effective decay
length $\xi _{eff}$

\begin{equation}
\tanh \frac{\xi ^{-1}}{\ell _{eff}^{-1}}=\frac{\xi ^{-1}}{\xi _{10}^{-1}\pm
\sqrt{\left( \ell _{soeff}^{-2}-\xi _{H}^{-2}\right) }}.  \label{ksieff}
\end{equation}

It is clearly seen that if the effective spin orbit interaction is so strong
that $\ell _{soeff}^{-1}\geq \xi _{H}^{-1}$, then the right hand side of (%
\ref{ksieff}) is real. Therefore in this case Eq. (\ref{ksieff}) provides us
by two solutions for $\xi _{1}^{-1},$ while $\xi _{2}^{-1}=0$. It is
necessary to mention that in the absence of ferromagnetic ordering $(H=0)$
due to degeneracy in spin orientation the critical current must not depend
on $\ell _{soeff}.$ In this situation only the root of equation
corresponding to the '+' sign in Eq. (\ref{ksieff}) should be considered
\begin{equation}
\tanh \frac{\xi _{11}^{-1}}{\ell _{eff}^{-1}}=\frac{\xi _{11}^{-1}}{\xi
_{0}^{-1}+\ell _{eff}^{-1}}  \label{smH1}
\end{equation}%
which provides the largest value of the decay length.

Solution of Eq. (\ref{ksieff})
\begin{equation}
\tanh \frac{\xi _{12}^{-1}}{\ell _{eff}^{-1}}=\frac{\xi _{12}^{-1}}{\xi
_{0}^{-1}+\ell _{eff}^{-1}+2\ell _{soeff}^{-1}}  \label{smH2}
\end{equation}%
with the smaller $\xi =\xi _{12}$ also exists at finite $H.$ (In the limit $%
H\rightarrow 0$ the prefactor before this exponential solutions goes to zero
\cite{Fuare} providing independence of the critical current on $\xi _{12}$).
At $\ell _{soeff}=\xi _{H}$ these two lengths, are equal to each other, $\xi
_{11}=\xi _{12}$. With further $H$ increase the right hand side of Eq.(\ref%
{ksieff}) becomes complex and Eq.(\ref{ksieff}) can be rewritten as
\begin{equation}
\tanh \frac{\xi ^{-1}}{\ell _{eff}^{-1}}=\frac{\xi ^{-1}}{\xi
_{10}^{-1}+i\xi _{20}^{-1}},\ \xi _{20}^{-1}=\sqrt{\xi _{H}^{-2}-\ell
_{soeff}^{-2}}.  \label{ksieffK}
\end{equation}

The sign '-' in Eq.(\ref{ksieff}) simply provides the equation for the
complex-conjugate solution of Eq.(\ref{ksieffK}).

In the limit $\ell _{eff}\ll \xi $ one can expand the hyperbolic tangent in (%
\ref{ksieffK}) in series keeping three first terms and get
\begin{equation}
\frac{\ell _{eff}}{\xi _{1}}=\sqrt{\frac{3\Gamma _{+}}{2}}\left[ 1+\frac{1}{%
10}\left( \frac{\ell _{eff}}{\xi _{01}}-1-\frac{\ell _{eff}^{2}}{\xi
_{20}^{2}\Gamma _{+}}\right) \right] ,  \label{ksi1dirty}
\end{equation}%
\begin{equation}
\frac{\ell _{eff}}{\xi _{2}}=\sqrt{\frac{3\Gamma _{-}}{2}}\left[ 1+\frac{1}{%
10}\left( \frac{\ell _{eff}}{\xi _{01}}-1+\Gamma _{+}\right) \right] ,
\label{ksi2dirty}
\end{equation}%
\begin{equation*}
\Gamma _{\pm }=\sqrt{\left( \frac{\ell _{eff}}{\xi _{01}}-1\right) ^{2}+%
\frac{\ell _{eff}^{2}}{\xi _{20}^{2}}}\pm \left( \frac{\ell _{eff}}{\xi _{01}%
}-1\right) .
\end{equation*}%
The expressions in the square brackets in (\ref{ksi1dirty}), (\ref{ksi2dirty}%
)\ give first order corrections to the dirty limit formula \cite{Fuare} for $%
\xi _{1}$ and $\xi _{2}.$ This approximation valid if%
\begin{equation}
\sqrt{\xi _{0}^{-2}+2\xi _{0}^{-1}\ell _{soeff}^{-1}+\xi _{H}^{-2}}\pm
\left( \xi _{0}^{-1}+\ell _{soeff}^{-1}\right) \ll \ell _{eff}^{-1}.
\label{valdirty}
\end{equation}

In the limit $\xi _{0},\ell _{soeff}\gg \xi _{H}$ the expression $\xi =\sqrt{%
\frac{D_{F}}{iH}\left( 1-\frac{2}{5}iH\tau \right) },$ $\tau
=l/v_{F},$\ follows from Eqs.(\ref{ksi1dirty}), (\ref{ksi2dirty}).
This formula was obtained before in Ref. \cite{Buzdin,Tagirov,Fominov} and can be
interpreted as a complex correction to the diffusion coefficient,
$D_{F}^{eff}=D_{F}\left( 1-\frac{2}{5}iH\tau \right) $ .

In the clean limit
\begin{equation}
A\gg \max \left\{ \ln \sqrt{A},~\ln \sqrt[4]{\frac{\ell _{eff}^{2}}{\xi
_{H}^{2}}-\frac{\ell _{eff}^{2}}{\ell _{soeff}^{2}}}\right\}
\label{condclean}
\end{equation}%
\begin{equation*}
A=1+\frac{\ell _{eff}}{\xi _{0}}+\frac{\ell _{eff}}{\ell _{soeff}}
\end{equation*}%
in the first approximation we may put the hyperbolic tangent in (\ref%
{ksieffK}) equal to unity and get%
\begin{equation}
\xi _{1}^{-1}=\xi _{10}^{-1},\quad \xi _{2}^{-1}=\xi _{20}^{-1}
\label{firstcl}
\end{equation}%
It is clearly seen that for $\ell _{soeff}^{-1}\rightarrow 0$ this formula
transforms into Eq.(\ref{ksi_c}). In the next approximation it is easy to
get that the corrections to (\ref{firstcl})
\begin{equation}
\xi _{1}^{-1}=\xi _{10}^{-1}-2p\exp \left( -\frac{2\ell _{eff}}{\xi _{10}}%
\right)   \label{ksi1clean}
\end{equation}

\begin{equation}
\xi _{2}^{-1}=\xi _{20}^{-1}+2q\exp \left( -\frac{2\ell _{eff}}{\xi _{10}}%
\right)  \label{ksi2clean}
\end{equation}%
where
\begin{equation*}
p=\xi _{20}^{-1}\sin \left( \frac{2\ell _{eff}}{\xi _{20}}\right) +\xi
_{10}^{-1}\cos \left( \frac{2\ell _{eff}}{\xi _{20}}\right)
\end{equation*}%
\begin{equation*}
q=\xi _{10}^{-1}\sin \left( \frac{2\ell _{eff}}{\xi _{20}}\right) -\xi
_{20}^{-1}\cos \left( \frac{2\ell _{eff}}{\xi _{20}}\right)
\end{equation*}%
are oscillating functions of $\xi _{H}.$

Eq.(\ref{ksieffK}) is equivalent to the system of equations for $\xi _{1}$
and $\xi _{2}$
\begin{equation}
\frac{\xi _{1}^{-1}}{\xi _{10}^{-1}}=\coth \frac{2\xi _{1}^{-1}}{\ell
_{eff}^{-1}}-a\cos \left( \frac{2\xi _{2}^{-1}}{\ell _{eff}^{-1}}-\arctan
\frac{\xi _{2}^{-1}}{\xi _{10}^{-1}}\right)  \label{sys1}
\end{equation}%
\begin{equation}
\frac{\xi _{2}^{-1}}{\xi _{20}^{-1}}=\coth \frac{2\xi _{1}^{-1}}{\ell
_{eff}^{-1}}-b\cos \left( \frac{2\xi _{2}^{-1}}{\ell _{eff}^{-1}}+\arctan
\frac{\xi _{1}^{-1}}{\xi _{20}^{-1}}\right)  \label{sys2}
\end{equation}%
\begin{equation*}
a=\frac{\sqrt{\xi _{10}^{-2}+\xi _{2}^{-2}}}{\xi _{10}^{-1}\sinh \frac{2\xi
_{1}^{-1}}{\ell _{eff}^{-1}}},\ b=\frac{\sqrt{\xi _{20}^{-2}+\xi _{1}^{-2}}}{%
\xi _{20}^{-1}\sinh \frac{2\xi _{1}^{-1}}{\ell _{eff}^{-1}}}
\end{equation*}

From the structure of equations (\ref{sys1}), (\ref{sys2}) it follows that
increase of $\xi _{20}^{-1}$ leads to increase of $\xi _{2}^{-1}.$ This, in
turn, results in increase of the second negative item in right hand side of (%
\ref{sys1}). Since $\xi _{1}^{-1}$ must be a positive value increase of $\xi
_{2}^{-1}$ should be accompanied by a jump, at a certain point, to the
positive branch of $\cos (x)$ leading to a discontinuity of $\xi
_{2}^{-1}(\xi _{20}^{-1})$ dependence. This consideration is proved by
numerical solution of (\ref{ksieffK}) (see Figs.1-3).

Figures 1 and 2 show the dependencies of $\xi _{1}^{-1}(\xi _{20}^{-1})$ and
$\xi _{2}^{-1}(\xi _{20}^{-1})-\xi _{20}^{-1}$ calculated for fixed values
of parameter $\xi _{10}^{-1}.$ Open triangles and circles in the figures
show the asymptotic dependencies (\ref{ksi1dirty}), (\ref{ksi2dirty}) and (%
\ref{ksi1clean}), (\ref{ksi2clean}), respectively. It is clearly seen that
in the parameter intervals $\xi _{20}^{-1}\leq 10\ell _{eff}^{-1},$ $\xi
_{10}^{-1}\geq 2\ell _{eff}^{-1},$ the expressions (\ref{ksi1clean}), (\ref%
{ksi2clean}) provide a good fit to the exact solution of equation (\ref%
{ksieffK}). The dirty limit formulas (\ref{ksi1dirty}), (\ref{ksi2dirty})
are valid up to $\xi _{20}^{-1}\leq 2\ell _{eff}^{-1}$ for $\xi
_{10}^{-1}\leq 2\ell _{eff}^{-1}.$ Figure 3 gives the ration of $\xi
_{2}/\xi _{1}$ as a function of $\xi _{20}^{-1}$ for a set of $\ell
_{eff}/\xi _{10}.$ At $H\rightarrow 0$ the oscillation length $\xi _{2}$
goes to infinity. Therefore the ratio is diverges at $\xi
_{20}^{-1}\rightarrow 0.$ With $H$ increase the ration rapidly decreases
approaching the law $\xi _{2}/\xi _{1}\propto \xi _{20}^{-1}$ at $\xi
_{20}^{-1}\geq 2.$

\begin{figure}[ht]
\centerline{\includegraphics[width=0.5\textwidth]{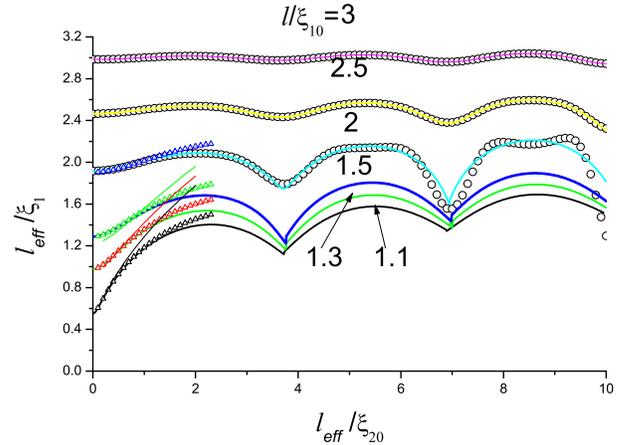}}
\caption{\label{fig:fig1} The decay length $\protect\xi _{1}^{-1}$ as a function of $\protect%
\xi _{20}^{-1}$ calculated for different values of
$\protect\xi_{10}^{-1}$.
The open circles are the asymptotic curves, which have been calculated from (%
\protect\ref{ksi1clean}) for $\protect\xi _{10}^{-1}=$2, 2.5 and 3. The open
triangles are the asymptotic curves calculated from (\protect\ref{ksi1dirty}%
) for $\protect\xi _{10}^{-1}=$1.1, 1.3, 1.5 and 2. The thin solid
lines are the the asymptotic dependencies following from Eq.
(\protect\ref{ksi1dirty}) without the correction in the square
brackets. These curves are calculated for $\protect\xi
_{10}^{-1}=$1.1, 1.3 and 1.5.}
\end{figure}

\begin{figure}[ht]
\centerline{\includegraphics[width=0.5\textwidth]{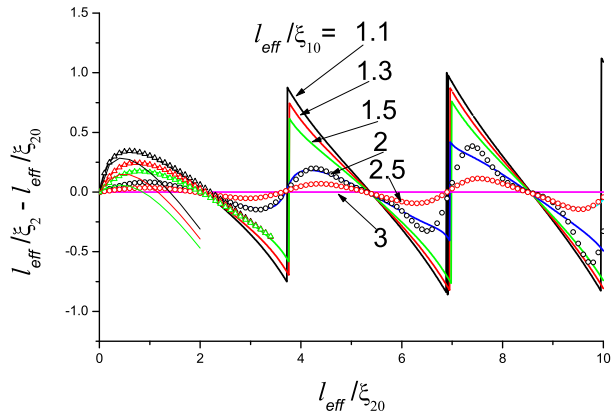}}
\caption{\label{fig:fig2} The difference between decay length $\protect\xi _{2}^{-1}$ and $%
\protect\xi _{20}^{-1}$ as a function of $\protect\xi _{20}^{-1}$
calculated for different values of $\protect\xi _{10}^{-1}$ shown in
figure. The open circles are the asymptotic curves, which have been
calculated from ( \protect\ref{ksi2clean}) for $\protect\xi
_{10}^{-1}=$2, 2.5 and 3. The open
triangles are the asymptotic curves calculated from (\protect\ref{ksi2dirty}%
) for $\protect\xi _{10}^{-1}=$1.1, 1.3, 1.5 and 2. The thin solid
lines are the the asymptotic dependencies following from Eq.
(\protect\ref{ksi2dirty}) without the correction in the square
brackets. These curves are calculated for for $\protect\xi
_{10}^{-1}=$1.1, 1.3 and 1.5.}
\end{figure}

\begin{figure}[ht]
\centerline{\includegraphics[width=0.5\textwidth]{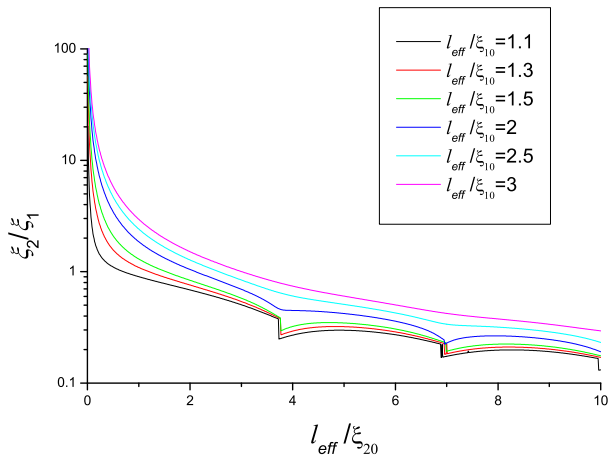}}
\caption{\label{fig:fig3} The ratio of oscillation and decay lengths
$\protect\xi _{2}/ \protect\xi _{1}$ as a function of $\protect\xi
_{20}^{-1}$ calculated for different values of $\protect\xi
_{10}^{-1}$. }
\end{figure}

The discovered behavior of $\xi _{2}$ and $\xi _{1}$ is quite general and
must be also observed in structures without ferromagnetic ordering. An
example is a normal filament of finite length, which is placed between
superconducting banks and is biased by a dc supercurrent. It was shown \cite%
{Kalenkov}, that the minigap induced to this filament from the S electrodes
is not a monotonous function of phase difference across the structure. This
behavior could be also explained in terms of specific dependencies of $\xi
_{2}$ and $\xi _{1}$ upon electron mean free path in current biased systems.

In summary, by solving the linearized Eilenberger equations we have
calculated the real, $\xi _{1},$ and the imaginary, $\xi _{2},$ part of a
decay length as a function of exchange energy $H$ and the mean free paths $%
\ell ,$ $\ell _{so},\ell _{z}$ and $\ell _{x}$ for ordinary, spin-orbit and
spin-flip electronic scattering in a ferromagnet. These parameters, $\xi
_{1},$ and $\xi _{2},$ characterize penetration of superconducting
correlations into a ferromagnet due to proximity effect and determine the
decay and oscillation lengths of a critical current in long SFS Josephson
structures. We have found the range of validity of the expressions (\ref%
{ksi_d}), (\ref{ksi_c}), which are commonly used for interpretation of
experimental data. In particular, the dirty limit expressions (\ref{ksi_d})
are valid if $\xi _{20}^{-1}\leq 0.5\ell _{eff}^{-1}$ for $\xi
_{10}^{-1}\leq 0.5\ell _{eff}^{-1}$. The corrected expressions (\ref%
{ksi1dirty}), (\ref{ksi2dirty}) can be used in a broader range of $\xi
_{20}^{-1}$ $\leq 2\ell _{eff}^{-1}$ and $\xi _{10}^{-1}\leq 2\ell
_{eff}^{-1}.$ Further increase of exchange field makes the the length $\xi
_{2}$ smaller than $\ell _{eff}$ thus breaking down the validity of
approximations used in derivation of the Usadel equations. It is interesting
to note that in certain parameter range the jumps occur in the dependence of
$\xi _{2}$ vs $\xi _{20}$, while $\xi _{1}$ remains a continuous function of
$\xi _{20}.$

We have also demonstrated that the intuitive knowledge about the relation
between $\xi _{1}$ and $\xi_2$, based on the dirty limit theory, has very
limited range of applicability and can not be used for $\xi _{H}$ $>5\ell $
or for $H\tau >0.1$. In particular, an increase of $H$ is not always
accompanied by a decrease of $\xi _{1}$ and in a certain parameter range $%
\xi _{1}$ may even increase with $H$. The fact that one may combine
reasonably large decay length with the smaller period of oscillations looks
rather attractive for possible applications of SFS Josephson junctions.

The authors thank A. D. Zaikin, L. Tagirov and A. I. Buzdin for
useful discussions. This work was supported in part by PI-Shift
Programme, RFBR grant 06-02-90865 and NanoNed programme under
project TCS 7029.

\end{document}